%% file: murs-2016.tex
\documentclass[10pt, conference, compsocconf]{IEEEtran}
\ifCLASSINFOpdf
\else
\fi

\usepackage[ruled,linesnumbered,lined]{algorithm2e}

\usepackage{graphicx}
\usepackage{subfigure}

\usepackage{hyperref}
\usepackage{multirow}
\usepackage{balance}
\usepackage{comment}
\usepackage{tikz}
\usepackage[numbers,sort&compress]{natbib}

\usepackage{color}
\definecolor{dkgreen}{rgb}{0,0.6,0}
\definecolor{gray}{rgb}{0.5,0.5,0.5}
\definecolor{mauve}{rgb}{0.58,0,0.82}
\begin{document}
%
\title{MURS: Mitigating Memory Pressure in Service-oriented Data Processing Systems}


\author{\IEEEauthorblockN{Xuanhua Shi\IEEEauthorrefmark{1},
Xiong Zhang\IEEEauthorrefmark{1},
Ligang He\IEEEauthorrefmark{2},
Hai Jin\IEEEauthorrefmark{1}, 
Zhixiang Ke\IEEEauthorrefmark{1},
Song Wu\IEEEauthorrefmark{1}}
\IEEEauthorblockA{\IEEEauthorrefmark{1}SCTS/CGCL, School of Computer Science and Technology\\
Huazhong University of Science and Technology, Wuhan, China}
\IEEEauthorblockA{\IEEEauthorrefmark{2}Department of Computer Science, University of Warwick, UK\\
Email: \{xhshi, wxzhang, hjin, zhxke, wusong\}@hust.edu.cn, ligang.he@warwick.ac.uk}
}


%


\maketitle

\begin{abstract}

Although a data processing system often works as a batch processing system, many enterprises deploy such a system as a service, which we call the service-oriented data processing system. It has been shown that in-memory data processing systems suffer from serious memory pressure. 
The situation becomes even worse for the service-oriented data processing systems due to various reasons. 
For example, in a service-oriented system, multiple submitted tasks are launched at the same time and executed in the same context in the resources, comparing with the batch processing mode where the tasks are processed one by one. 
Therefore, the memory pressure will affect all submitted tasks, including the tasks that only incur the light memory pressure when they are run alone. 
In this paper, we find that the reason why memory pressure arises is because the running tasks produce massive long-living data objects in the limited memory space. 
Our studies further reveal that the long-living data objects are generated by the API functions that are invoked by the in-memory processing frameworks. 
Based on these findings, we propose a method to classify the API functions based on the memory usage rate. Further, we design a scheduler called MURS to mitigate the memory pressure. 
We implement MURS in Spark and conduct the experiments to evaluate the performance of MURS. 
The results show that when comparing to Spark, MURS can 1) decrease the execution time of the submitted jobs by up to 65.8\%, 2) mitigate the memory pressure in the server by decreasing the garbage collection time by up to 81\%, and 3) reduce the data spilling, and hence disk I/O, by approximately 90\%. 

\end{abstract}


%
\IEEEpeerreviewmaketitle

\input{intro}
\input{motivation}

\input{mur}

\input{scheduler}

\input{impl}
\input{eval}
\input{related}
\input{concl}

\bibliographystyle{IEEEtran}
{\scriptsize
\bibliography{ref}
}

\end{document}

%% file: intro.tex
\section{Introduction}

Many popular distributed data processing systems are sped up by in-memory computing models. These systems, such as Spark~\cite{zaharia2012resilient} and Flink~\cite{hueske2012opening}, are usually developed in Java, Scala and other managed languages. In-memory data are stored as data objects in memory, which can bloat the memory as shown in the literature~\cite{bu:bloat}. Limited memory space used by the submitted jobs will result in the memory pressure. Although these managed languages provide garbage collection(GC) to reclaim the useless data objects, caching data in memory may worsen the memory pressure as the data objects usually have a long lifetime in memory~\cite{lulu:deca}. 
Data processing systems often work as batch processing systems. Jobs are submitted to systems individually and processed without the users' intervention. Many enterprises also deploy a data processing system  as a service, such as Spark SQL~\cite{armbrust2015spark} and Hive~\cite{ashish:hive}, which we call the service-oriented data processing system. The situation of memory pressure becomes even worse in service-oriented data processing systems for the following reasons. 

First, in a service-oriented data processing system, multiple tasks are launched and executed at the same time in the resources (such as in a cluster), comparing with the batch processing mode where the tasks are processed one by one.
Second, in the service-oriented system, multiple jobs are executed in the same context. In addition to sharing the memory and CPU cores, the service-oriented systems prefer caching the sharing data in memory in order to speed up all related jobs.
Finally, some services are even oversold to the tenants as they take the chance that all tenants may not submit their jobs at the same time.  

The above features of the service-oriented data processing systems cause the heavy memory pressure, which affects all jobs, although some jobs only incurs the light memory pressure if they are run alone in the batch processing mode.


Hot keys and large intermediate results are two common causes of the heavy memory pressure, which consequently harms the performance of data processing systems~\cite{fang2015interruptible}. Hot keys may result in an out-of-memory error and cause the system crash. Large intermediate results, such as Java collection, may lead to frequent garbage collections. When the garbage collection time becomes a big proportion of the entire execution time, the performance of the data processing system is significantly degraded. If the situation persists, the system may even crash. Most works mitigate the memory pressure by customizing the memory managers or tuning the applications in batch processing~\cite{www:spark-tuning, nguyen2015facade, fang2015interruptible, lulu:deca, nguyen:yak}. However, these works may not be effective in service-oriented data processing systems due to their unique features we discussed above. In this work, we propose a new method to address this difficult issue. In the method, we classify the tasks in terms of memory pressure. A task that causes the heavy memory pressure is called a \textit{heavy task}, while the task with the light impact on memory pressure called a \textit{light task}. Based on the classification, we suspend the heavy tasks so as to enable the light tasks to complete quickly. By doing so, the resources are released from running heavy tasks and the memory pressure is reduced for all tasks.


The memory pressure originates from massive long-living data objects, which are produced by the running tasks in the limited memory space. Our studies further show that the root cause for generating these long-living data objects is the API functions called by the tasks. These API functions are based on key-value pairs~\cite{dean2008mapreduce, zaharia2012resilient, hueske2012opening, isard2007dryad}. We find that the memory space used by the function APIs can be traced. The memory usage of the function APIs can be classified into four coarse-grained models: constant, sub-linear, linear and super-linear. Each memory usage model has a different influence on the memory pressure. The tasks with the linear or super-linear memory models, which are the tasks that process more input dataset, are more likely to be the heavy tasks. These four models are combined independently in each task with a strict order. We propose the \textit{Memory Usage Rate} to determine which model a task belongs to. Further, we design a Memory-Usage-Rate based Scheduler, called MURS, which can efficiently mitigate heavy memory pressure and avoid memory overflow. The scheduler first collects the memory usage rate of the currently running tasks, and then selects the heavy tasks and suspend them. After the heavy tasks are suspended, the light tasks can then complete quickly because of the light memory pressure.

We implement MURS in the service-oriented Spark and conduct the extensive experiments. The experimental results show that the execution time of the jobs in the service-oriented data processing system can be reduced by up to 65.8\% (2.9X). The garbage collection, which directly measures the memory pressure, can decrease by a maximum of 81\%. MURS can also help improve the scalability of the service-oriented system, as the system equipped with MURS can still provide high quality of service even if the memory is in shortage. 

In summary, this work offers the following three main contributions.

\begin{itemize}

\item We analyze the memory utilization of various data processing systems, and find that the memory pressure originates from the function APIs called by the data processing systems during the task executions. The invocation of the API functions generate massive long living data objects, which consume a large amount of memory. 


\item Based on the above findings, we build four memory usage models to measure the impact a task imposes on memory pressure. The proposed models are independent of the data processing systems (and their function APIs). We also propose to use memory usage rate to identify which model a task belongs to. We design the memory usage rate based scheduler called MURS, which releases the memory pressure substantially.


\item We implement a prototype system of MURS based on Spark. The proposed method can be ported to other similar service-oriented data processing systems. The experimental results have shown that our system can significantly improve system performance, reduce garbage collection and improve the scalability of the service-oriented system.

\end{itemize}

The rest of the paper is organized as follows. Section II introduces the motivation of our work. Section III describes the four types of memory usage rate. Section IV presents the scheduling mechanism for the tasks with different memory usage rates. Section V presents the implementation of MURS. Performance evaluation is carried out in Section VI to show the effectiveness of our scheduler. In Section VII, we discuss the related work. Finally, Section VIII concludes this paper.

%% file: motivation.tex
\section{Motivation}
\label{sec:motivation}

Memory is a critical type of resource in current data processing systems, especially in the in-memory computing systems~\cite{shi:mammoth}. Limited memory space leads to memory pressure and can cause frequent garbage collection or out-of-memory errors~\cite{fang2015interruptible}, both of which seriously affect the performance of the data processing system.

\begin{figure}[!t]
\centering
\includegraphics[width=0.35\textwidth]{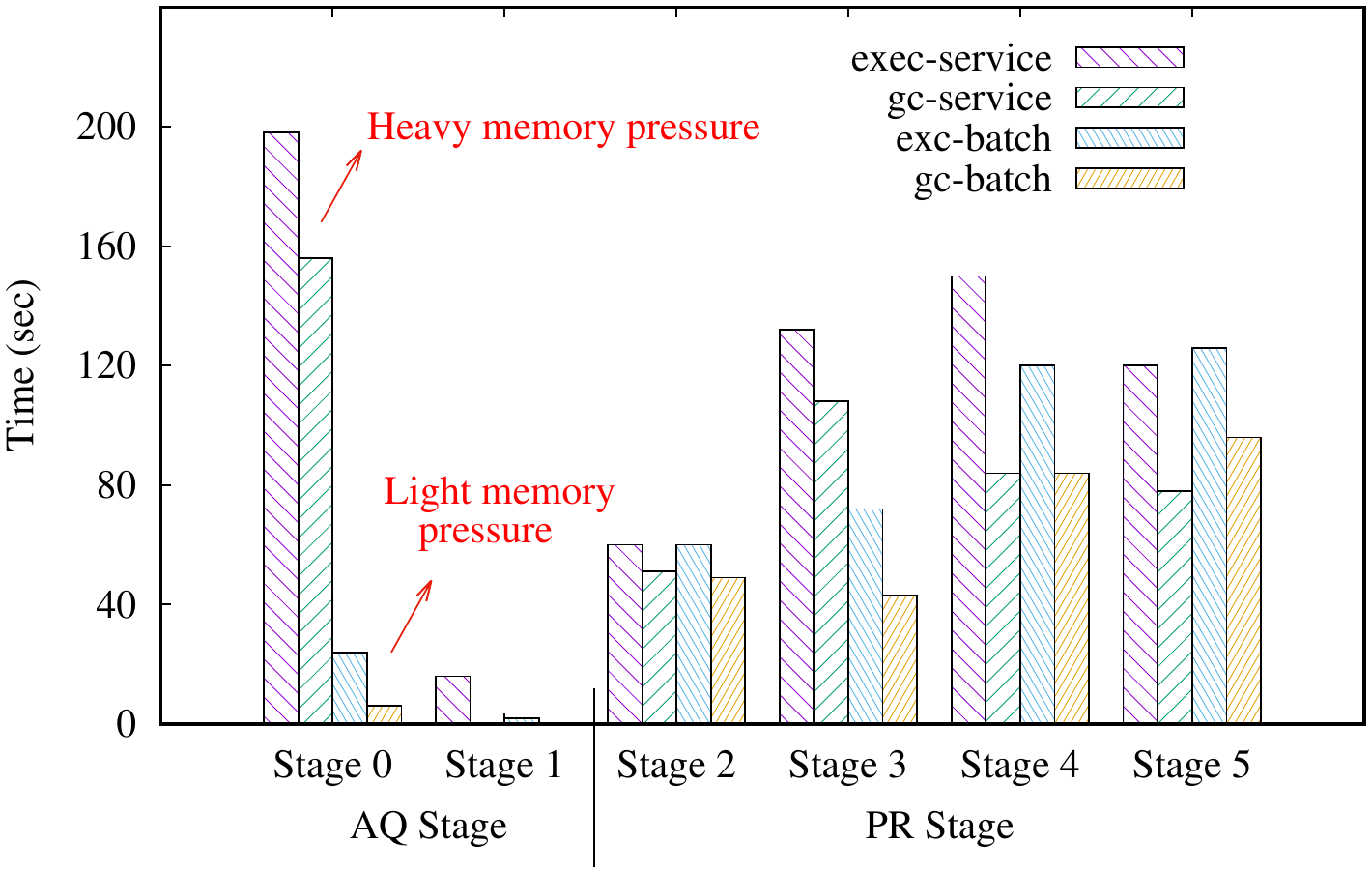}
\vspace{-2mm}
\caption{AQ suffers memory pressure from PR}
\vspace{-6mm}
\label{fig:memorypressure}
\end{figure}

A data processing system is often deployed as a batch processing system, but can also be deployed as a service and work as a service-oriented system. To understand the impact of memory pressure on the service-oriented data processing systems and identify the cause of inefficiency, we investigate different applications in Spark. Spark can also work as a service-oriented system through the Spark Job Server. We choose two types of applications, PageRank(PR)  
is a typical benchmark in Spark, and Aggregation Query(AQ) is the common big data benchmark in SQL~\cite{www:benchmark}. In order to represent a general situation, a semantic-identical hand-written Spark program is used for AQ. The input datasets of PR and AQ is webbase-2001 (30GB) and HiBench Random Writer (50GB), respectively. We first run the tasks in the service-oriented mode, in which we submit PR and AQ simultaneously to the service-oriented Spark and run them with a fair scheduler in Spark. As a comparison, we also run PR and AQ in the batch mode, in which PR and AQ are processed one after the other. We record the execution times and garbage collection times of all tasks under these two modes, which are plotted in Figure~\ref{fig:memorypressure}. We then analyze the  memory pressure through the results.   

We observe that PR caches intermediate data in memory and iteratively compute the result. One iteration corresponds to one stage in Spark. Because the caching data is alive as long as the job itself, the memory space becomes less gradually
and the task computation may suffer. If this occurs, it indicates the memory pressure caused by PR is heavy. Comparing with PR, however, AQ only contains some simple operations. AQ has only two processing stages and during its execution, a very small amount of intermediate shuffle data are generated (although the input data of AQ is larger than that of PR). Furthermore, both the execution time and garbage collection time are low in AQ, and the memory pressure is much lighter than PR. The result of PR in the batch mode also verifies its high memory pressure and heavy garbage collection, which are the results labelled as exec-batch and gc-batch in Figure~\ref{fig:memorypressure}. As we can see, the garbage collection time accounts for a very large proportion of the execution time.

When multiple jobs, such as PR and AQ, are submitted to and run by the Spark service simultaneously, the jobs are run with a fair scheduler provided by Spark. Although AQ has much lighter memory pressure than PR, all running tasks suffer from the heavy memory pressure produced by PR. We find that the execution time (exec-service in Figure~\ref{fig:memorypressure}) of every task in each stage of PR has little change, except some maximum execution times. This is because almost all memory pressure comes from PR and therefore the executions of PR in the service mode and the batch mode show similar trends. However, The execution of AQ in the service mode is very different from that in the batch mode. This is because in the service mode, both applications are run simultaneously and therefore AQ suffers from the memory pressure produced by PR, even if AQ is a light task itself. In the batch mode, since the applications are run one after another. The high memory pressure created by PR will not affect the running of AQ.

In summary, our results implicate that in a service-oriented system, 1) the heavy memory pressure will result in frequent garbage collection, which consumes most of the time and reduces the throughput; 2) the light tasks suffer from heavy memory pressure produced by the heavy tasks; and 3) the heavy tasks obtain the resources later because these resources are occupied chronically by light tasks, and the heavy tasks themselves are the source of memory pressure.

By observing the first stage of PR and AQ, we discover that the tasks of PR and AQ invoke different function APIs, which determine the impact of each task on memory pressure. If we can identify and classify these tasks by the characteristics of the function APIs, we can suspend the heavy tasks and leave adequate memory space to the light tasks when the memory pressure shows up. This can improve the throughput of the service-oriented systems and allow all tasks to run with enough memory space and hence light memory pressure.

%% file: mur.tex
\section{Memory Usage Model in Tasks}



As discussed in previous section, some tasks consume less memory while some use much more memory as they produce massive long living data object, which are mainly generated by function APIs in the processing pipeline. Although there are various function APIs, some of them manifest a similar characteristic in terms of memory usage. Based on this observation, we build the models to capture the memory usage characteristic of a function API, and then the memory usage rate is used to determine which model the task belongs to.


\subsection{Memory usage models of APIs}

\begin{table*}[!t]
\small
\centering
\caption{Function APIs in Distributed Data Processing System} 
\begin{tabular}{ c | c | c | c | c | c | c }

\hline
\multirow{2}{*}{\textbf{Community}} & \multicolumn{2}{|c|}{ \multirow{2}{*}{\textbf{Core API} }} & \multirow{2}{*}{\textbf{Application Systems}} & \multicolumn{3}{|c}{\textbf{Partial Function APIs}} \\
\cline{5-7}
 & \multicolumn{2}{|c|}{} & & constant & sub-linear & linear \\
\hline
Hadoop & MapRedcue~\cite{vavilapalli2013apache} & Crunch & Pig, Hive, Yarn & map & reduce & \\
\hline
Microsoft & Drayd~\cite{isard2007dryad} & DryadLINQ & Scope, MadLINQ & map & reduce & join \\
\hline
Spark & \multicolumn{2}{|c|}{RDD~\cite{zaharia2012resilient}} & Spark SQL~\cite{armbrust2015spark}, GraphX~\cite{xin2013graphx} & map & reduceByKey & groupByKey \\
\hline
Flink & \multicolumn{2}{|c|}{Dataset~\cite{www:flink}} & Table~\cite{www:flink}, Gelly~\cite{www:gelly} & where & distinct & join \\
\hline
Google & MapReduce & FlumeJava~\cite{flumejava} & Tenzing, Pregel, Sibyl & parallelDo & combinValue & groupByKey \\
\hline

\hline
\end{tabular}
\vspace{-2mm}
\label{table:apps}
\end{table*}

A data processing system provides several function APIs, which can be used to implement various applications. These function APIs take as input the input dataset and produce another dataset. The type of data in the dataset may be different. Table~\ref{table:apps} lists the function APIs provided by popular data processing systems. Most function APIs are sourced from MapReduce, a famous computing framework. Other function APIs are used to control the execution of jobs, the typical case of which is the shuffle operations. The shuffle operations are used between the \textit{Map} and the \textit{Reduce} phase, and are regarded as the separation point of a job, because the tasks after the shuffle operations (i.e., \textit{Reduce} tasks) need all results calculated by the tasks before the shuffle operations (\textit{Map} tasks).

These function APIs are all based on key-value pairs: (\textit{K}, \textit{V}). Some function APIs omit the parameter \textit{K} or \textit{V} for the convenience of users. Rather, the default value of the omitted parameters are used during the processes of \textit{map} or \textit{reduce}. The memory space is used by a function API to store the living data objects, because temporary data objects will be reclaimed by the garbage collection. The memory demand determines the characteristic of memory usage. Based on the key-value pairs, the memory size of living data objects is related to both \textit{K} and \textit{V} in the following ways.

\begin{itemize}

\item If the function API does not distinguish the parameter \textit{K}, it processes a record without involving other records. A record will produce a new record accordingly. If the new record is cached in memory, the consumed memory size will certainly increase. If the new record is processed as the input of the next function API or write-to-disk, it will be regarded as a temporary data object and quickly transmitted to the next function API. The size of temporary data object is ignored as it will be reclaimed in the next round of garbage collection.

\item If the function API distinguishes \textit{K}, it will involve all records to process these records with a particular \textit{K}. The function APIs that involve all records are usually called shuffle. While the shuffling stage processes the records to get all \textit{V} with a particular \textit{K}, two operations can be performed on \textit{V}: aggregation and non-aggregation.

\item If the function API does not aggregate the \textit{V}, it only puts \textit{V} in a collection without involving other \textit{V}. The collection contains the intermediate data and has a long lifetime because it is alive until the task processes all records. The collection is usually called the shuffle buffer. After a record is processed, the size of the collection increases by one element and thus the memory size of the shuffle buffer must increase.

\item If the function API aggregates the \textit{V}, the intermediate collection will be replaced by the aggregated value. We also call these data objects with long lifetime the shuffle buffer. Aggregation of \textit{V} means some operations will be performed on all \textit{V} with a particular \textit{K} and produce a new value. Thus, the resulting memory size of the collection will increase when \textit{K} has not appeared yet.

\end{itemize}

\begin{figure}[!t]
\centering
\includegraphics[width=0.3\textwidth]{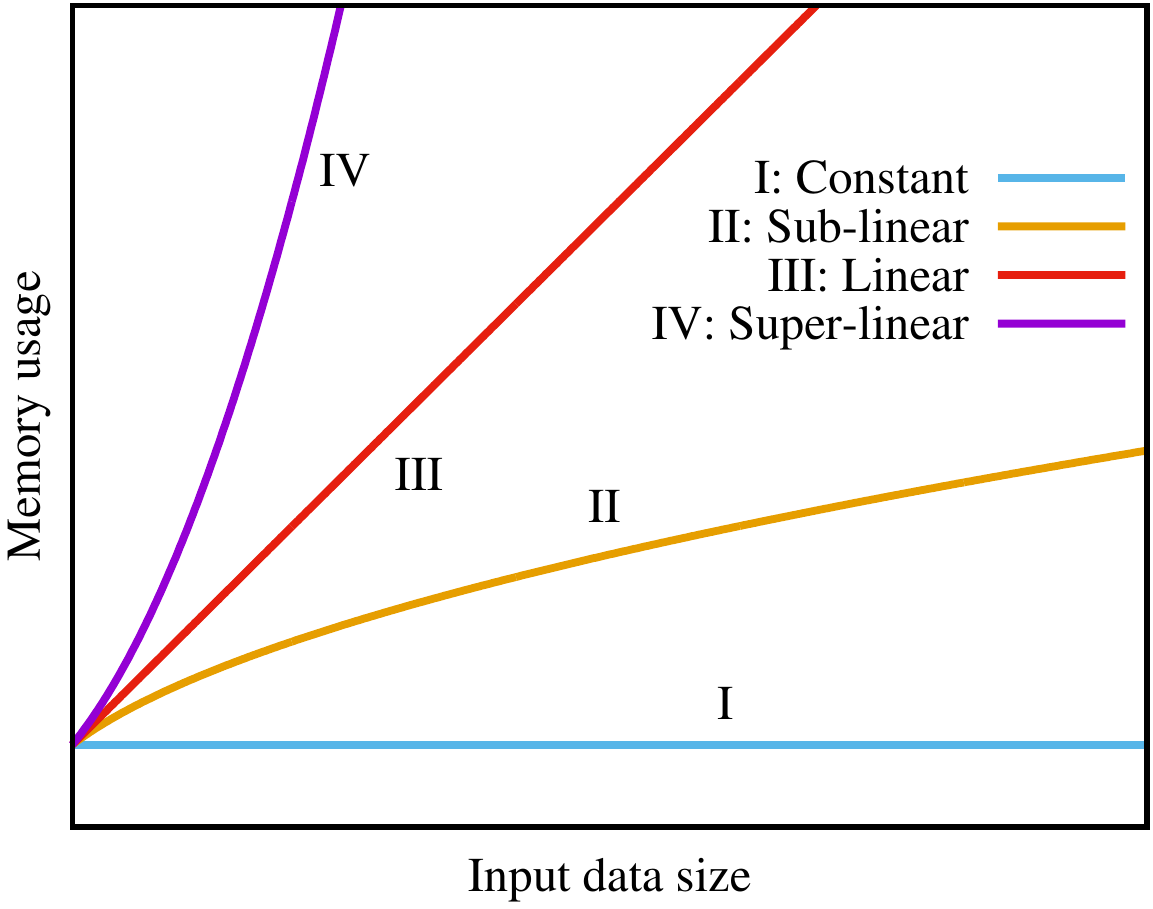}
\vspace{-2mm}
\caption{Four coarse-grained models of function API}
\vspace{-4mm}
\label{fig:mur}
\end{figure}

As the operations on \textit{K} and \textit{V} decide the size of the living data objects in memory, we build four models in this work to measure the memory usage of each function API when it processes a unit of data. The models are based on the size of the records that are processed, not on the number of the processed records, because the size of a record in each dataset is different. Memory usage refers to the memory space used to store the data objects with long lifetime except the garbage data objects. The four models are shown in Figure~\ref{fig:mur}. We determine the memory usage model of a function API by using the following lemmas.

\newtheorem{lemma}{Lemma}
\begin{lemma}[Constant] The memory usage model can be defined as constant (Line I in Figure~\ref{fig:mur}) only when the following conditions are true:
\begin{enumerate}
\item The function API does not distinguish the key, \textit{K};
\item The resulting data will not be cached in memory.
\end{enumerate}
\end{lemma}

\begin{lemma}[Sub-Linear] The memory usage model can be defined as sub-linear (Line II in Figure~\ref{fig:mur}) only when the following conditions are true:
\begin{enumerate}
\item The function API distinguishes the key, \textit{K};
\item The function API aggregates the value, \textit{V};
\item \textit{K} appears randomly in the input dataset.
\end{enumerate}
\end{lemma}

Note that the reason why we require that \textit{K} appears randomly is because the size of intermediate data will increase only when the \textit{K} has appeared. If most \textit{K} gathers around some neighbouring records, the size of intermediate data will increase linearly.

\begin{lemma}[Linear] The memory usage model can be defined as linear (Line III in Figure~\ref{fig:mur}) when the following conditions hold:
\begin{enumerate}
\item The function API distinguishes \textit{K};
\item The function API does not aggregate \textit{V}.
\end{enumerate}
\end{lemma}

Both cache operations and the appearance pattern of \textit{K} affect the  size of intermediate data. Thus although some function APIs have the same operations on \textit{K} and \textit{V}, their memory usage models may be different. The constant model requires that intermediate data are not cached in memory. The increasing model is determined by the size of the cached data objects. The sub-linear model also requires the random appearance of \textit{K}. When a function API does not satisfy the above conditions, we need to redefine their models. In other words, the model is defined not only by the function APIs, but also by the user-defined function or data distribution. The memory usage model should be redefined when the following conditions are true:

\begin{itemize}

\item If the function API does not distinguish the \textit{K} and the result data are cached in memory, the speed of increasing size in memory can be 1) \textit{linear} (Line III in Figure~\ref{fig:mur}) when the function API does not work based on formal result; 2) \textit{super-linear} (Line IV in Figure~\ref{fig:mur}) when the function API produces larger result data, such as computing a histogram of the appeared numbers and all their divisors; 3) \textit{sub-linear} (Line II in Figure~\ref{fig:mur}) when the function API produces smaller result data along with the computation. 

\item If a function API i) distinguishes \textit{K}, ii) does not aggregate \textit{V}, iii) \textit{K} has not appeared yet in the input dataset or the appearance of \textit{K} is not random. The memory usage model of the function API should be linear.

\end{itemize}

Note that when function APIs belong to the same linear model, they can also be distinguished because the slope of the line in Figure~\ref{fig:mur} may be different. Steeper the slope is, the heavier impact the API function has on memory pressure. Based on the slope and four models, we can distinguish all function APIs in various data processing systems.

\subsection{Memory usage models in tasks}
\label{subsec:taskmodel}

A task is implemented by at least one function API. Some systems define only one function API in each task, such as Hadoop. Other systems define the function APIs in a task according to the shuffle operations in the user-defined program, such as Spark and Dryad. As the shuffle operations are used to split the jobs, they implement both shuffle write and shuffle read. Thus, we consider the memory usage of a task in three phases: read, process, and write. The read and write phase of a task only contain one shuffle function API, or no function API if they read/write from/to disk or print in screen. If a task has a process phase, the phase contains several function APIs. 

The read and write phases of a task have independent memory usage models with a strict order. However, the memory usage model of the process phase is different. A function API in the process phase is always of the constant model. They never distinguish  \textit{K} and do not need to calculate all intermediate data. Thus they process each record as a temporary data object and quickly pass the intermediate data to the next function API. All constant models will not be shown in the memory usage of a task if the write phase of the task has the shuffle operations. This is because the size of the shuffle buffer is much bigger than the constant models. However, when the task caches intermediate data in memory, these intermediate data will be transmitted to the next function API after completing the current function API. Under this circumstance, the model is redefined. The redefined model will be combined with the independent models in a task. When there are several memory usage models in a task, we can only monitor the current memory usage model when we schedule the task. When reducing the current memory pressure, we use the current memory usage model to calculate the memory usage of the task.

Based on the memory usage models in a task, we only need the slope of each line in Figure~\ref{fig:mur} to determine the current memory usage model. We term the slope the \textit{memory usage rate}. Thus, the memory usage rate of a task is defined as the memory size of the newly produced long-living data objects when a task processes a unit of input data.

%% file: scheduler.tex
\section{Design of MURS}
\label{sec:desgin}

Memory pressure essentially describes the usage of heap in managed languages. In MURS, we first compute the heavy tasks which may have the linear or super-linear models, or have a large input dataset. The fundamental scheduling idea of MURS is to suspend the heavy tasks as the memory pressure occurs, and resume the suspended tasks when the memory pressure recedes or light tasks are completed.

Firstly, the memory pressure occurs when the proportion of the used heap has reached the threshold value. The threshold is set based on the triggering condition of garbage collection. In addition, we set two thresholds in the scheduler of MURS. The first threshold, called the yellow value, is used to indicate the memory pressure. When the percentage of the long living data objects in the heap meets the yellow value, it suggests the frequent full GC will occur. Full GC means that the garbage collector will clean all the heap, which is usually computationally expensive. Another threshold value, called as red value, is used to avoid spilling. The red value represents the level of memory pressure under which the out-of-memory error will occur or suggests that some data will be spilled to the disk. The default yellow and red values are 0.4 and 0.8, respectively, based on our evaluation. When the memory pressure is high, we will reduce the value.

With the yellow and red values, an accurate percentage of long living data objects in the heap actually determines the efficiency of the threshold. JVM splits the heap to young generation and old generation. Minor GC cleans the young generation and moves living data objects to the old generation. Thus the percentage of the heap usage after a minor GC represents the living data objects in the heap. After each full GC, dead data objects in old generation will also be reclaimed and then we revise the percentage of long living data objects in the heap according to the current percentage of heap usage.



\IncMargin{0.4em}
\SetAlFnt{\small}
\begin{algorithm}[!t]

\SetKwInOut{Input}{Input}\SetKwInOut{Output}{Output}
\SetKwProg{Fn}{Function}{}{end}
\SetKwFunction{CST}{ComputeSuspendTasks}
\SetKwFunction{CS}{ComputeSpill}
\SetKwData{F}{\small{final}}

\Input{Array of running tasks $R$\;}
\Output{Array of suspended tasks $S$\;}
get the Memory Usage Sampler $Sampler$\;
get the Memory Manger $SM$ of System\;
get the Memory Manger $JM$ of JVM\;
get the queue including suspended tasks $SQ$\;
\lIf{Usage of $JM$ is lower than the yellow value}{return}
\lElseIf{Usage of $JM$ is lower than the red value}{\CS}
\lElseIf{$SQ$ is not empty}{return}
\lElse{\CST}

\BlankLine
\Fn{\CST}{
  $freeMemory \leftarrow JM.freeMemory$\;
  $consumption[] \leftarrow SM.tasksMemoryConsumption$\;
  $rate[] \leftarrow Sampler.getMemoryUsageRate$\;
  $percent[] \leftarrow Sampler.getCompletePercent$\;
  $S \leftarrow R$\;
  \While{$freeMemory > 0$}{
    $minRateTaskId \leftarrow reate[].min$\;   
    \lIf{\CS}{reduce the running cores and return}
    $memoryNecessary \leftarrow comsumption[taskId] * (1-percent[taskId])$\;
    $freeMemory -= memoryNecessary$ \;
    $S$ remove minRateTaskId \;
    push minRateTaskId to $SQ$\;
    $rate[]$ remove minRateTaskId \;
  }
  \KwRet{$S$}
}
\BlankLine
\Fn{\CS}{
  $taskId \leftarrow$ task Id \;
  $totalMemory \leftarrow JM.totalMemory$ \;
  \lIf{$comsumption[taskId] > totalMemory / R.length$}{\KwRet{True}}
  \Else{
      $memoryNecessary \leftarrow comsumption[taskId] / percent[taskId]$\;
      \lIf{$memoryNecessary > totalMemory / R.length$}{\KwRet{True}}
      \lElse{\KwRet{False}}
      }
}
\caption{The scheduling mechanism on JVM}
\label{code:scheduler}
\end{algorithm}

The indicators of memory pressure serve as the basis of the memory usage rate based scheduler. The scheduling strategy is presented in Algorithm~\ref{code:scheduler}. The input of the scheduler is the current running tasks and the output is the proposed suspended tasks. The scheduler records the indicators of memory pressure. A \textit{Sampler} is designed to record the real-time values of the metrics regarding the currently running tasks. The Sampler runs seasonally to update the values of the metrics for each task and compute the current memory usage rate, such as the size of input dataset, the number of processed records, and the size of the results.

If the memory pressure is light or there are already suspended tasks, we return immediately without proposing suspended tasks (lines 4, 6). If the memory pressure has reached the red value, MURS will try to avoid spilling (line 5). When the memory usage of the JVM heap reaches the yellow value, we obtain the memory usage rates of all running tasks from the sampler and other details of current memory usage (lines 9-12). We use the measures to process the currently running tasks in the following order of memory usage model: constant, sub-linear, linear, and super-linear. The available memory is then the free memory subtracting the memory required for the currently running tasks. When the free memory is not enough for running the remaining unprocessed tasks, the scheduler will stop processing tasks (lines 14-20). This method is able to distinguish the tasks that have the sub-linear model but are run as the heavy tasks when processing the larger datasets, or the tasks that have the super-linear model but are run as the light tasks when they consume less memory. The processed tasks are removed and the remaining tasks will be returned. The returned tasks are heavy tasks and will be suspended. In order to avoid potential starvation, the suspended tasks are pushed to a queue (line 21). The FIFO algorithm allows us to resume the first suspended task to avoid starvation. 

MURS suspends the proposed heavy tasks to prevent memory pressure from increasing at a high rate. Note that if all running tasks are heavy tasks, MURS still schedules the tasks with the same algorithm. It always selects the tasks that have relatively lower memory usage rate out of all tasks. 

Function \textit{ComputeSpill} is designed to avoid spilling. As the running tasks share the JVM heap, the maximum memory space allowed for each task must be less than $M/N$ ($M$ is the memory size and $N$ is the number of running threads). When the actual memory consumption of a task exceeds the maximum space, the spill or out-of-memory error will occur. The running tasks that satisfy the condition of \textit{ComputeSpill} (lines 27, 30) are also suspended to reduce the degree of parallelism in order to acquire adequate memory space after the memory pressure is released.   

When a running task is completed, we resume the execution of a suspended task popped out of the queue. After the memory pressure recedes, reflected by the phenomenon that the usage of JVM heap decreases to below the yellow value after a full GC, the remaining suspended tasks will also be resumed. Namely, the completion of a running task only causes one suspended task to be resumed while the memory usage below yellow value will cause all suspended tasks to be resumed.

%% file: impl.tex
\section{Implementation}

We implement MURS in Spark 1.6.0 with approximately 1000 Scala codes. MURS contains a scheduler and a sampler. The scheduler implements the Algorithm~\ref{code:scheduler}. 
The sampler works seasonally to collect and compute all metrics in the scheduler. 
The sampler updates the following metrics in a task after a record is processed:

\begin{enumerate}

\item Read phase. 
The updated metrics include the number of processed records, the number of total records, the size of total records and the size of the shuffle buffer.

\item Process phase. 
The updated metric is the size of cached data objects unrolled by this task.

\item Write phase. 
The updated metrics are the size of the shuffle buffer, the number of write records, and the number of total records. 

\end{enumerate}

The sampler also updates some metrics of the memory mangers, including the current memory usage of a task, free memory space, and the current proportion of used heap. Based on these metrics, current memory usage rate of a task is computed as the quotient of two increments: $\bigtriangleup size_{used\_memory} / \bigtriangleup size_{processed\_records}$. All computed values of the memory usage rate are stored in a buffer. The changing trend of the memory usage rate determines the model that the task belongs to.

The scheduler detects the total memory usage from the sampler seasonally. When it reaches the yellow value in Algorithm~\ref{code:scheduler}, the scheduler proposes the suspended tasks. All tasks process the data in {\ttfamily \small InterruptibleIterator}, which can be interrupted by the system or the scheduler. We use a flag to guide the suspension of the {\ttfamily \small InterruptibleIterator}. After MURS proposes the suspended tasks, we update the flag of each task in the task scheduler. If the flag is true, the {\ttfamily \small InterruptibleIterator} will suspend itself and the task is also suspended. When MURS resumes the task, it only needs to set the flag of this task to be false.


When we deploy the service-oriented Spark, we choose Spark Job Server 0.6.2~\cite{www:jobserver}. Each job in Spark runs within the only {\ttfamily \small SparkContext}. The Spark Job Server first defines the shared context. We set the scheduler mode as FAIR and build a scheduler pool in the context. 
All jobs are submitted to the Spark Job Server instead of Spark. The Spark Job Server will resubmit the job to the shared context in Spark.

%% file: eval.tex
\section{Evaluation}

\subsection{Configuration}

We conducted the extensive experiments to evaluate the performance of the methods proposed in this work. In the experiments, we use four nodes as the workers and one node as the master. Each node has two eight-core Xeon-2670 CPUs and 64GB memory. The file system is mounted on a SAS disk, running the RedHat Enterprise Linux 5 (kernel 2.6.18). The JDK version is 1.7.0 for Spark 1.6 and Spark Job Server 0.6.2. We record not only the execution time of each application, but also the detailed runtime situations.

We choose four applications to evaluate the performance: Scan Query (SQ) and Aggregation Query (AQ) which are exploratory SQL queries in ~\cite{www:benchmark}, Sort and  PageRank(PR) which are typical benchmark applications in Spark. For each query, a semantic-identical hand-written Spark program are used for general cases. PR is a typical iterative computations, one of its most important features is that it will cache the data in memory. Another important features is that each iteration of PR is similar to the Join Query in SQL queries.
For Sort and AQ, the datasets are produced by HiBench Random Writer with 1B unique key numbers. The sizes of the input datasets for Sort and AQ are 30GB and 50GB, respectively. For SQ and PR, we use the real graphs, webbase-2001 (30GB)~\cite{boldi:webgraph}, to evaluate the performance of MURS. The key-value pairs of each record in all input dataset have the similar size. We use the heap size to evaluate different memory pressure. The garbage collection time is used to measure the memory pressure. These applications are grouped to evaluate different scenes and submitted to the Spark Job Server together.

\subsection{Memory pressure without caching}

Most current data processing systems are designed based on MapReduce. But only a part of them provide the in-memory computing model that caches the data in memory to speed up the system. Thus, we first evaluate these applications without caching the data in memory, which represents the common frameworks that work with the key-value pairs, such as Hadoop and Hive.

\begin{figure*}[!t]
\centering
\subfigure[Sort+SQ]{
\label{fig:subfig:sort-grep}
\includegraphics[width=0.23\textwidth]{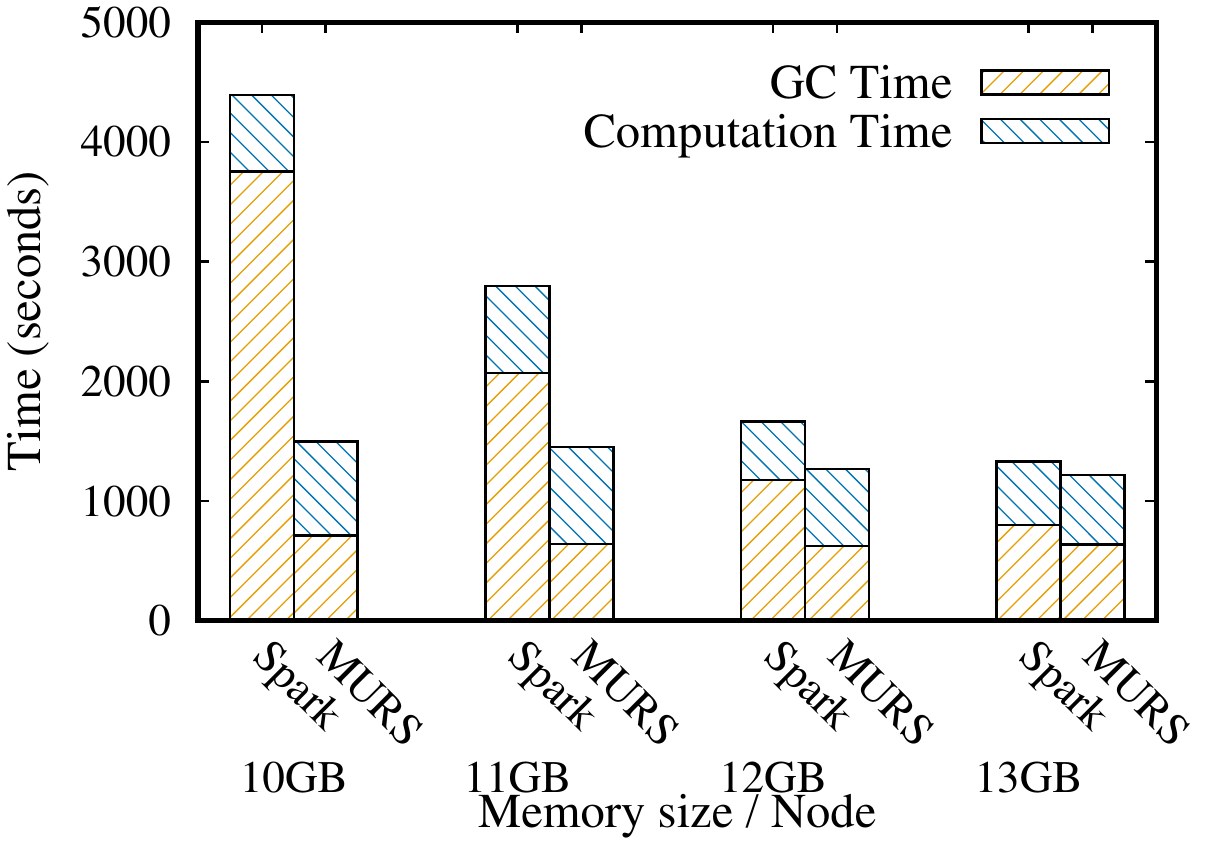}}
\subfigure[AQ+SQ]{
\label{fig:subfig:wc-grep}
\includegraphics[width=0.23\textwidth]{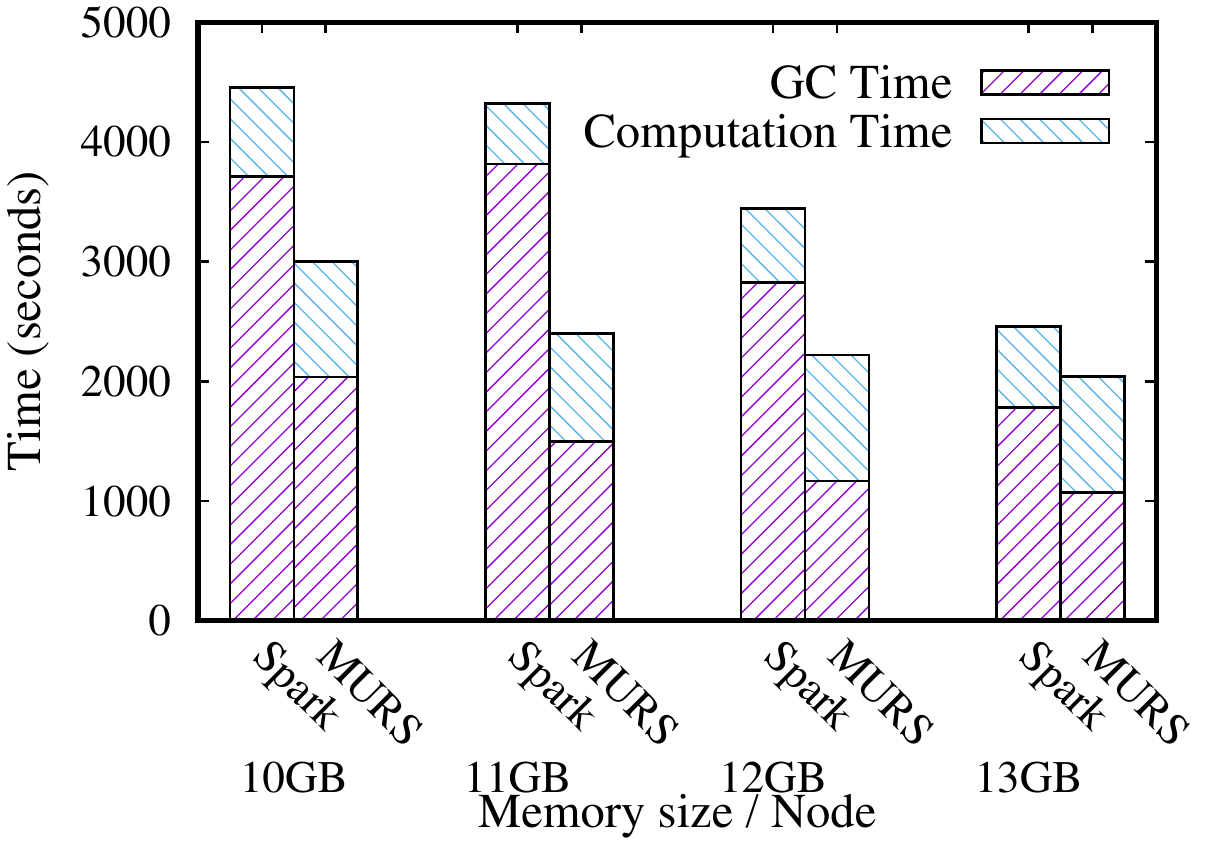}}
\subfigure[Sort+AQ+SQ]{
\label{fig:subfig:sort-wc-grep}
\includegraphics[width=0.23\textwidth]{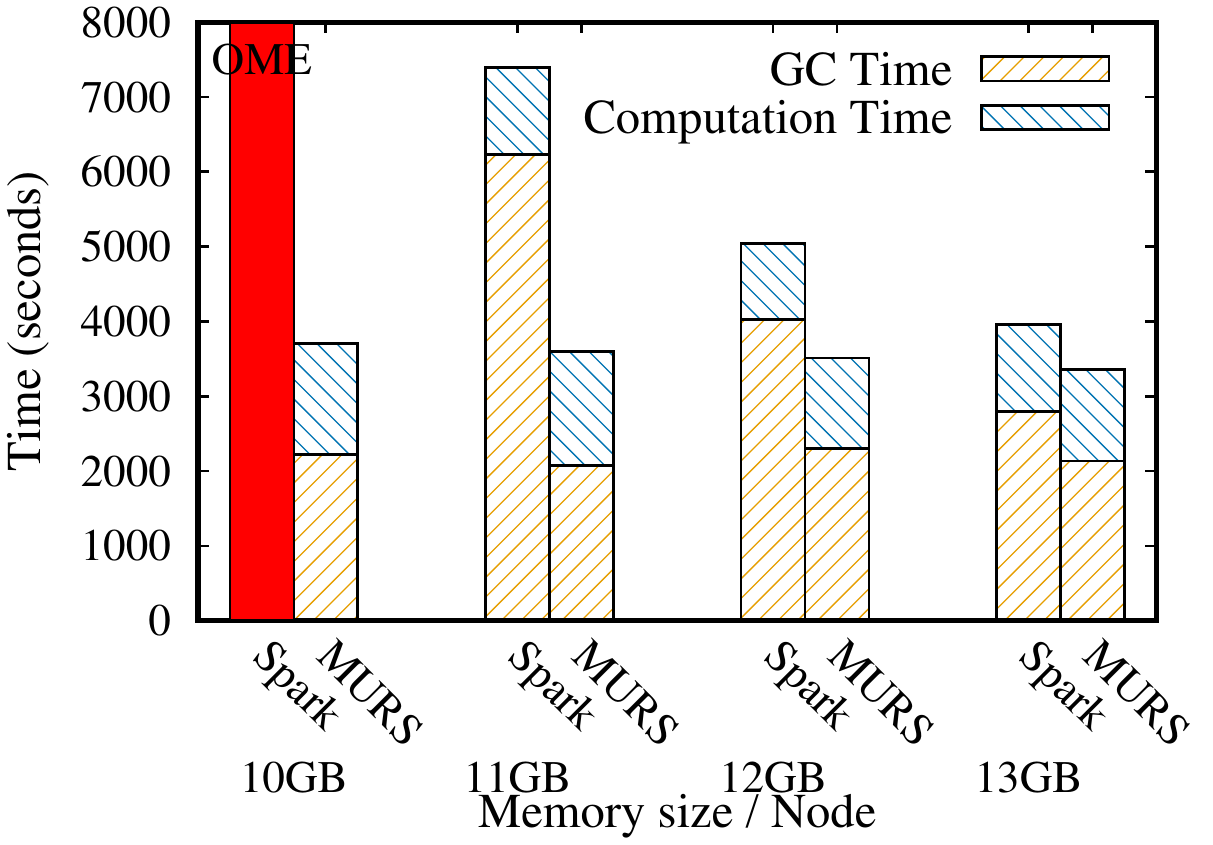}}
\subfigure[Active tasks]{
\label{fig:subfig:activetasks}
\includegraphics[width=0.23\textwidth]{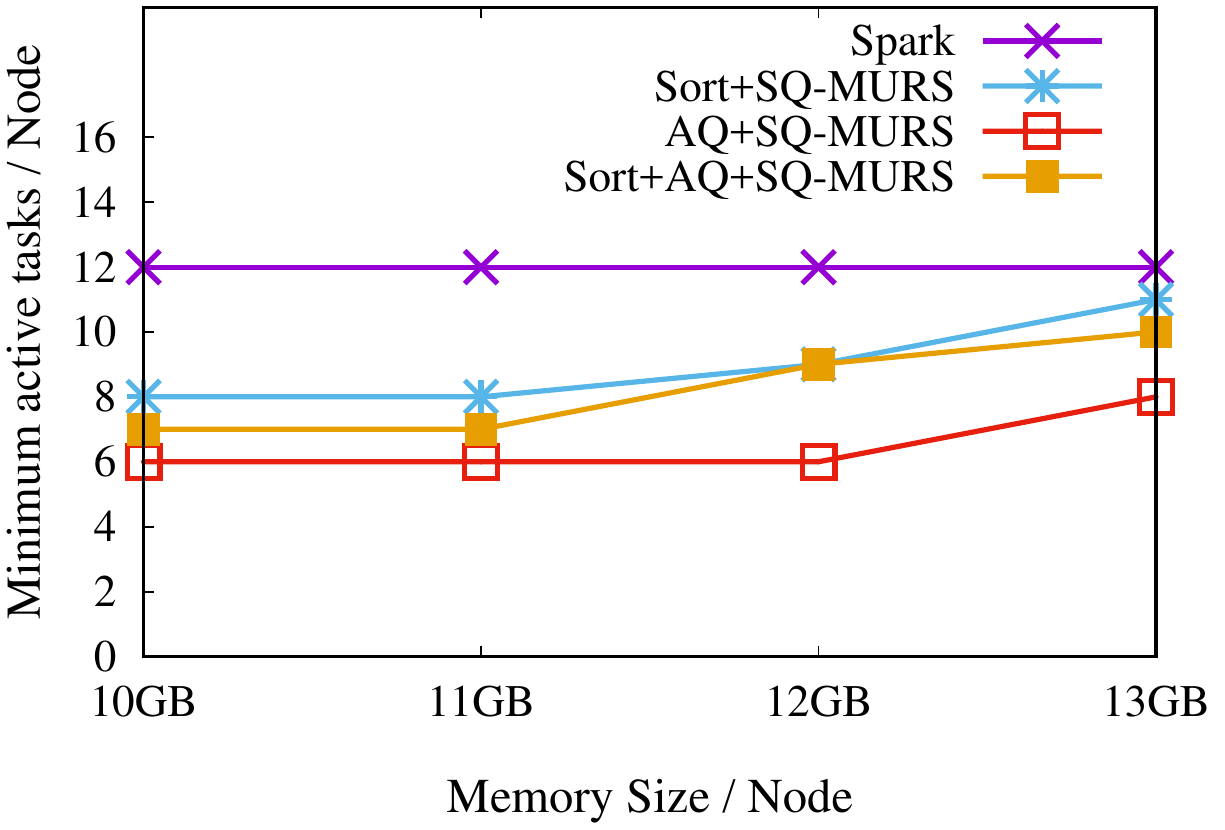}}
\vspace*{-2mm}
\caption{Results of memory pressure without caching}
\vspace*{-4mm}
\label{fig:pressurewithoutcache}
\end{figure*}

We choose three applications: Sort, AQ, and SQ. These applications do not perform the caching operations. Each application has similar implementation in MapReduce. SQ reads the data from the disk and filters these records which satisfy the given conditions. Most data objects are temporary. The shuffle buffers in Sort and AQ contain the data objects with long lifetime. Thus the tasks in Sort and AQ are the light to heavy tasks in MURS. The results of each submission are shown in Figure~\ref{fig:pressurewithoutcache}. The best improvement achieved by MURS is 1.8x to 2.9x compared to Spark in each evaluation. The reduction of the garbage collection contributes most to the improvement.

When two applications are submitted to the server, MURS works better in Sort and SQ. Comparing Sort to AQ, the heavy memory pressure occurs in the different phase. The \textit{sort} operation is implemented in the \textit{read phase} of the tasks. Thus the tasks in the last stage of Sort suffer from the heavy memory pressure during their read phases. However, the \textit{reduce} operation in AQ is realized in the \textit{write phase} of the tasks. The heavy memory pressure is caused by the write phase of the tasks in the first stage. Moreover, massive temporary data objects are produced by the function API \textit{flatMap} before the write phase. They may be currently alive in the heap during the write phase in AQ. 
Thus, MURS suspends more tasks in AQ as the heap size in AQ is occupied by the temporary data objects, as shown in Figure~\ref{fig:subfig:activetasks}. These temporary data objects result in frequenter garbage collections. 

When three applications are submitted, out-of-memory (OME) is thrown by the server. Although the live shuffle buffer in this case is smaller than the case with two applications, some other data objects occupy much heap space, such as the recording of living applications, the handler of disk writing, and the global configurations of Spark. Spark provides the spilling capability to overcome the shortage of memory. However, it cannot completely avoid the OME.

MURS mitigates the memory pressure in each application and we find that the performance decreases slowly as the heap size decreases. The result shows that the cost of garbage collection is stable. As MURS suspends the heavy tasks, the slow increase in execution time is due to the increase in computation time.  

\subsection{Memory pressure with caching}

Some frameworks provide the caching mechanism for in-memory computation, such as Spark and Flink. Although the in-memory data caching speeds up the execution of a job, some works~\cite{bu:bloat, nguyen2015facade} show that it results in greater memory pressure because the cached data live as long as the job, especially when they are deployed as service-oriented systems. 

\begin{figure}[!t]
\centering
\includegraphics[width=0.4\textwidth]{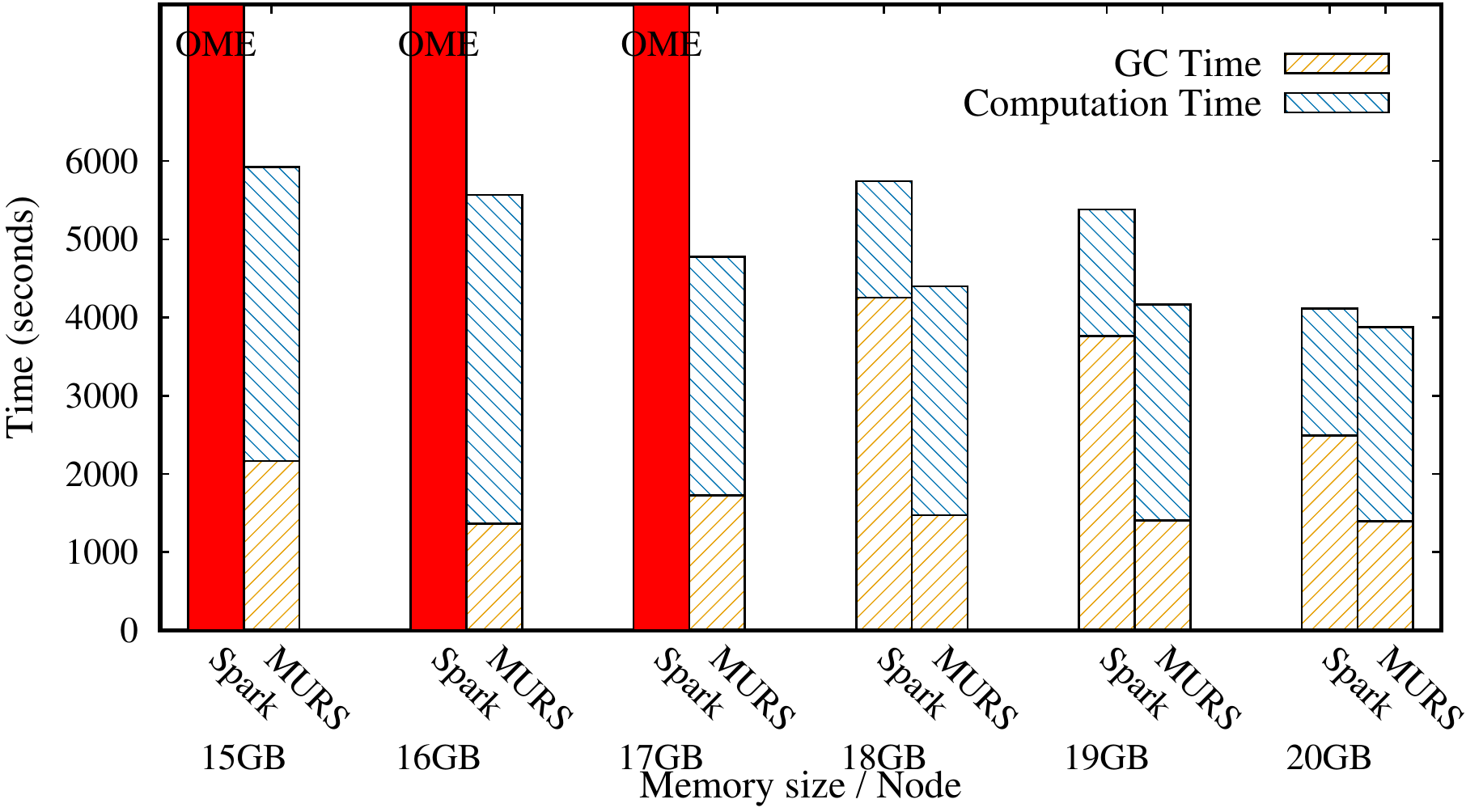}
\vspace{-2mm}
\caption{The Execution and GC Time of PR and AQ in Server}
\vspace{-2mm}
\label{fig:cache-total}
\end{figure}

PR and AQ are used as the experimental applications. PR is an iterative application. In the experiments, we run 5 iterations of the application. PR caches the intermediate data in memory after the first iteration. AQ is submitted at the second iteration. 
We adjust the heap size to show the performance of MURS, while the input dataset is 30GB, as shown in Figure~\ref{fig:cache-total}. When the heap size is less than 17GB, the service-oriented Spark throws the Out-of-Memory error (OME). MURS can provide the service when the heap size is reduced to 15GB. While they are both working, MURS improves the performance by up to 23.4\%, and the memory pressure is reduced by 65.4\%.

\begin{figure}[!t]
\centering
\includegraphics[width=0.4\textwidth]{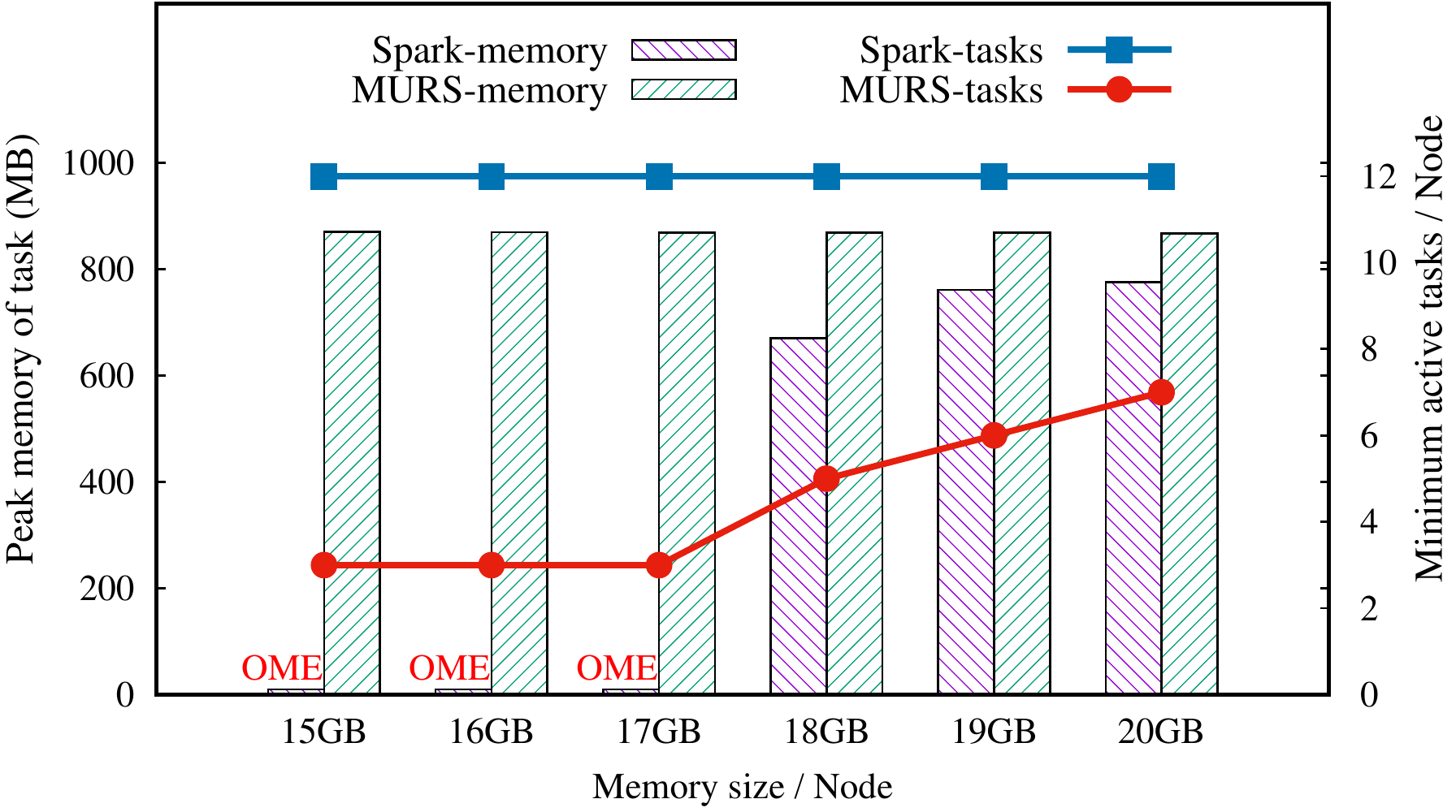}
\vspace{-2mm}
\caption{The peak usage memory of tasks and minimum active tasks}
\vspace{-4mm}
\label{fig:cache-peak}
\end{figure}

Less heap size in each node means more memory pressure in the server. When the heap is exhausted, tasks will throw the OME. With the caching data in memory, The service-oriented Spark throws the OME as the heap size of each node is 17GB. However, MURS is able to mitigate the heavy memory pressure and still performs well when the heap size of each node is reduced to 15GB. MURS suspends heavy tasks and  thus the remaining light tasks can utilize more memory. Figure~\ref{fig:cache-peak} shows that the peak memory usage of all tasks in MURS is consistently larger than that in Spark. The number of active tasks in MURS is reduced to ensure there is the memory available for the running tasks. In other words, MURS has better scalability than the service-oriented Spark.

\begin{figure}[!t]
\centering
\includegraphics[width=0.3\textwidth]{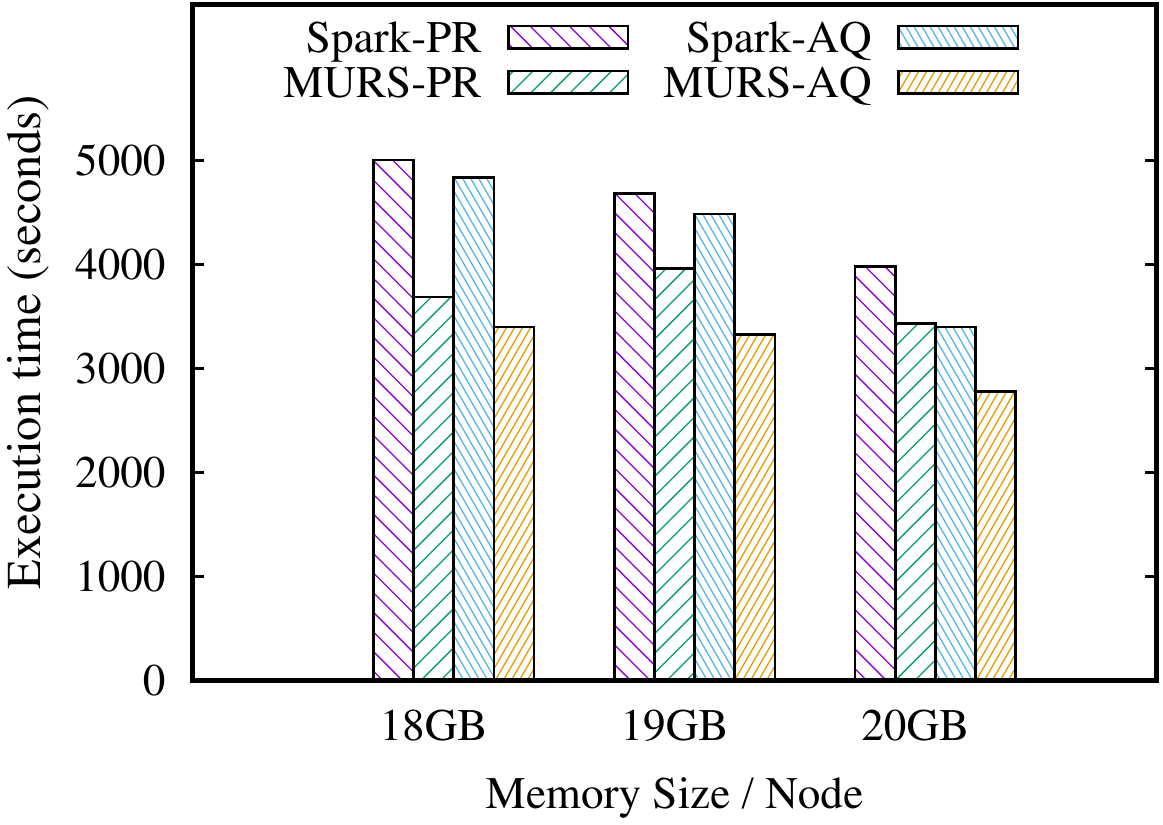}
\vspace{-2mm}
\caption{The execution time of each application}
\vspace{-2mm}
\label{fig:cache-prwc}
\end{figure}

\textbf{Potential Starvation} The mitigation strategy of memory pressure delays the computation of suspended tasks. This may cause the starvation of the suspended tasks. MURS address this issue in the application level. 
When PR and AQ are submitted to the server, t
he execution time of each application is shown in Figure~\ref{fig:cache-prwc}. The result shows there is no starvation for PR. Benefiting from MURS, the performance of PR is improved by up to 24.4\%. The performance of AQ, which contains light tasks, also achieves a 29.8\% improvement. 

MURS applies the FIFO algorithm when it resumes the suspended tasks. This avoids the long wait of these tasks. Otherwise, suspending these tasks is going to mitigate the memory pressure not only for the currently running light tasks, but also the heavy tasks running next. In the application level, the transient suspension and light memory pressure are actually the trade-off made by MURS. 


\textbf{Avoiding the spill} MURS sets the alarming threshold to avoid the spill when the memory pressure is heavy in the systems that allow the spill. 
Table~\ref{table:spill} shows the spill tasks in PR and AQ. There are no the spill tasks in AQ in MURS, while the spill tasks in PR decrease from 32\% to 2.5\%.
We notice that the error exists when avoiding the spill. The sampler in MURS counts two important metrics for a task: the percentage of the processed records in the total records, and the currently allocated memory space for this task. We can quickly obtain the required memory space based on these two metrics. When the memory pressure reaches the alarming threshold, we suspend a part of the running tasks and leave enough memory space for the remaining running tasks. As the estimate is based on sampling, the error exists in terms of the estimated size of the allocated memory space, especially when the value in some key-value pairs is a collection. 
Thus, there are even fewer spilling tasks in MURS that the figures reported in the experiments.

\begin{table}[!t]
\small
\centering
\caption{Spill Tasks in MURS and Spark}
\begin{tabular}{ c | c | c | c | c | c }

\hline
\multirow{2}{*}{\textbf{App}} & \multirow{2}{*}{\textbf{Tasks}} & \multicolumn{2}{| c |}{\textbf{Spark}} & \multicolumn{2}{| c }{\textbf{MURS}} \\
\cline{3-6}
 & & spill & percent & spill & percent \\
\hline
AQ & 1000 & 91 & 9\% & 0 & 0  \\
\hline
PR & 1500 & 480 & 32\% & 37 & 2.5\% \\
\hline

\hline
\end{tabular}
 
\vspace{-4mm}
\label{table:spill}
\end{table}

%% file: related.tex
\section{Related Work}

The problem of memory pressure in a system has been studied for several years. Different methods have been proposed to address this issue, such as tuning of garbage collection, memory management and task scheduling. Most of these works focus on the offline calculation in data processing systems. 

\textbf{Garbage Collection} Most methods of garbage collection tuning are based on the features of different applications, such as Spark applications~\cite{www:spark-tuning}, Cassandra applications~\cite{www:cassandra}. These methods typically take the following two approaches: 1) replacing other garbage collection algorithms, such as Concurrent Mark Sweep (CMS) or Garbage First (G1), 2) tuning the important parameters, such as the ratio of young generation and old generation, to avoid frequent GC activity and data copying. Most researches attempt to rebuild the algorithm of garbage collection for common applications. For example, Yang et al.\cite{yang:fullgc} describes an incremental query model for reference calculations to optimize the cause of full GC. This garbage collection algorithm is universally useful to most memory pressure cases. Rodrigo et al.\cite{rodigo:NGeneration} proposed a N-Generational garbage collector for big data memory management, which also reduces the data copying in garbage collection. Tuning garbage collection requires the deep understanding and proficient skills on the data processing systems. 

\textbf{Memory Management} The memory analysis of data-intensive big data applications are put forward by Bu et al.~\cite{bu:bloat}. They consider the bloat of memory in object-oriented languages and design the bloat-aware paradigm: 1) merging small data to big data; 2) manipulating the data by directly accessing the buffers. Some memory management techniques at the system level also extend the language-level optimization, such as region based memory management (RBMM)~\cite{nguyen2015facade, nguyen:yak} and lifetime based memory management~\cite{lulu:deca}. Nguyen et al.~\cite{nguyen2015facade} advise the users to mark the class information in the applications to decompose the data object to regions. Then the memory can be allocated or reclaimed by the regions. Based on the same theory, they also proposed a new garbage collector to mange the data objects and control the objects separately to extend the method to iterative computations~\cite{nguyen:yak}. Lu et al.~\cite{lulu:deca} propose the lifetime-based memory management. They decompose the data objects based on the data container, which decides the lifetime of data objects. The decomposed data objects will impose less pressure on the memory and can be allocated and reclaimed by regions. Memory management provides an effective way to manage the memory pressure and has been investigated by a large body of research.

\textbf{Task Scheduler} As the memory pressure is caused by the running tasks, the task scheduler can also play a role in releasing the memory pressure or improving the efficiency of memory usage. Fang et al.~\cite{fang2015interruptible} design the interruptible tasks for data-parallel programs. They classify the memory consumption of a task into four parts: local data structures, processed input, unprocessed input and result. Based on their novel programming model, suspending the interruptible tasks can release parts of the memory consumption of random tasks when memory pressure mounts. The suspended tasks can be resumed when memory pressure decreases. Interruptible tasks can be resumed with the remaining in-memory data. Pu et al.~\cite{pu2016fairride} implement a new policy called FairRide, which  is able to fairly allocate the memory cache to multiple users with the shared files through the efficient blocking. This policy is the first to satisfy all three desirable properties: isolation-guarantee, strategy-proofness and Pareto-efficiency. Based on different scheduling standards, the task scheduler always adapts to different demands and is easy to be implemented.

%% file: concl.tex
\section{Conclusion}

In this work, we aim to mitigate the memory pressure in current service-oriented data processing systems. We analyze the memory usage of various function APIs enclosed in these systems, and build three coarse-grained models to classify the function APIs. Further, we propose to use the memory usage rate as a uniform criteria to measure the impact of a task on memory pressure in the service-oriented systems. Based on the memory usage rate, we develop a scheduler called MURS. MURS can suspend the tasks that cause heave memory pressure, and speed up the tasks with light memory pressure. MURS can be implemented in most data processing systems running in the service-oriented context. We conduct the extensive experiments to evaluate the effectiveness of the proposed methods and MURS. The results show that comparing with Spark, MURS can reduce the execution time of tasks by up to 65.8\%, mitigate the memory pressure by up to 81\% and avoid the task spill by approximately 90\%.